\documentclass[apj,numberedappendix]{emulateapj}
\usepackage{adjustbox}
\usepackage{natbib}
\usepackage{placeins}
\usepackage{apjfonts}
\usepackage{hyperref}

\usepackage{xcolor}

\begin{document}

\title{Exploring the SDSS Photometric Galaxies with Clustering Redshifts}
\shortauthors{Rahman et al.}

\author{
Mubdi Rahman\altaffilmark{1}, 
Alexander J. Mendez\altaffilmark{1},
Brice M\'{e}nard\altaffilmark{1,2},\\ 
Ryan Scranton\altaffilmark{3},
Samuel J. Schmidt\altaffilmark{3},
Christopher B. Morrison\altaffilmark{4},
Tam\'{a}s Budav\'{a}ri\altaffilmark{5}
}

\altaffiltext{1}{Department of Physics \& Astronomy, Johns Hopkins University, 3400 N. Charles Street, Baltimore, MD 21218, USA}
\altaffiltext{2}{Kavli IPMU, the University of Tokyo, Kashiwa 277-8583, Japan}
\altaffiltext{3}{Department of Physics, University of California, One Shields Avenue, Davis, CA 95616, USA} 
\altaffiltext{4}{Argelander-Institut f\"ur Astronomie, University of Bonn, Auf dem H\"ugel 71, 53121 Bonn, Germany}
\altaffiltext{5}{Department of Applied Mathematics and Statistics, Johns Hopkins University, 3400 N. Charles Street, Baltimore, MD 21218, USA}
\email{mubdi@pha.jhu.edu}

\begin{abstract}
We apply clustering-based redshift inference to all extended sources from the Sloan Digital Sky Survey photometric catalogue, down to magnitude $r=22$. We map the relationships between colours and redshift, without assumption of the sources' spectral energy distributions (SED). We identify and locate star-forming, quiescent galaxies, and AGN, as well as colour changes due to spectral features, such as the $4000\,$\AA\ break, redshifting through specific filters. Our mapping is globally in good agreement with colour-redshift tracks computed with SED templates, but reveals informative differences, such as the need for a lower fraction of M-type stars in certain templates. We compare our clustering-redshift estimates to photometric redshifts and find these two independent estimators to be in good agreement at each limiting magnitude considered. Finally, we present the global clustering-redshift distribution of all Sloan extended sources, showing objects up to $z\sim0.8$. While the overall shape agrees with that inferred from photometric redshifts, the clustering redshift technique results in a smoother distribution, with no indication of structure in redshift space suggested by the photometric redshift estimates (likely artifacts imprinted by their spectroscopic training set). We also infer a higher fraction of high redshift objects. The mapping between the four observed colours and redshift can be used to estimate the redshift probability distribution function of individual galaxies. This work is an initial step towards producing a general mapping between redshift and all available observables in the photometric space, including brightness, size, concentration, and ellipticity.
\end{abstract}


\section{Introduction}
\label{sect:intro}

For the vast majority of extragalactic sources, spectroscopic information is unavailable. Globally, the fraction of objects with spectroscopic redshift measurements is currently of order one percent and is unlikely to increase in the foreseeable future. Consequently, for the vast majority of sources, redshift information is restricted to that which can be extracted from broad-band photometry. With such data sets, redshift inference relies on the ability to determine the \emph{mapping} between the photometric and redshift spaces. Currently, this mapping has been established in one of two ways: (i) the ``photometric'' redshift technique, which uses predicted relationships between colours and redshift obtained by convolving libraries of spectral energy distributions (SEDs) with the wavelength response of the system; and (ii) using information extracted from the clustering of matter, rather than the colours of the sources. This more recent approach is referred to as the clustering redshift technique. 

Classical photometric redshift inference using SED libraries has played a major role in extragalactic astronomy and will continue to provide valuable redshift information. However, it suffers from limitations in certain cases: first, it relies on the strong, and potentially dangerous, assumption that reference SED libraries fully represents all sources observed. Secondly, it typically uses only a subset of the photometric space, the measured colours, and does not include the remainder of the available information (including brightness, size, and concentration). Finally, it requires perfect knowledge of the response function of the system, including the filter and atmospheric transmission functions, which can vary with time and location on the sky.

Recently, clustering-based redshift inference has emerged as a new avenue to infer the redshift distribution of arbitrary samples of extragalactic sources, requiring only the on-sky ($\alpha$, $\delta$) position of the sources. Since it does not use the \emph{values} of measured colours to infer redshift, this approach has the potential to alleviate the limitations faced by the template-based photometric redshifts. The clustering-based redshift inference has been suggested and attempted for over three decades in various manners \citep{seldner79, Phillipps87, landy96, Schneider06, ho08, newman08, menard13, schmidt13, mcquinn13}. Over the past few years, our team has aimed at turning this idea into a usable tool for redshift estimation with real data \citep{menard13,schmidt13,schmidt15,rahman15b,rahman15a} and has successfully applied it to optical, near-infrared, and submillimeter observations.

In this paper, after briefly presenting the concept of clustering redshift inference (Section~\ref{sect:dataanalysis}), we apply this technique to the SDSS photometric galaxies which include over 100 million sources. Without any prior knowledge of source SEDs or the system's photometric response as a function of wavelength, we characterise the mapping between colours and redshift (Section~\ref{sect:colsep}), compare it to colour-redshift tracks obtained with galaxy and quasar spectral templates, compare our inferred distributions to photometric redshifts, and present our estimate of the redshift distribution of the SDSS survey as a function of limiting magnitude (Section~\ref{sect:photsep}).
This analysis is an initial step towards fully characterising the mapping between photometric space and redshift. This mapping can ultimately combine all the observables in the photometric space, such as brightness, size, concentration, and environment, exhausting all the available photometric information to infer redshifts.

\section{Data Analysis}
\label{sect:dataanalysis}

\subsection{The SDSS Photometric Galaxies}
\label{subsect:sdss}

\begin{figure}
\includegraphics[scale=1.0]{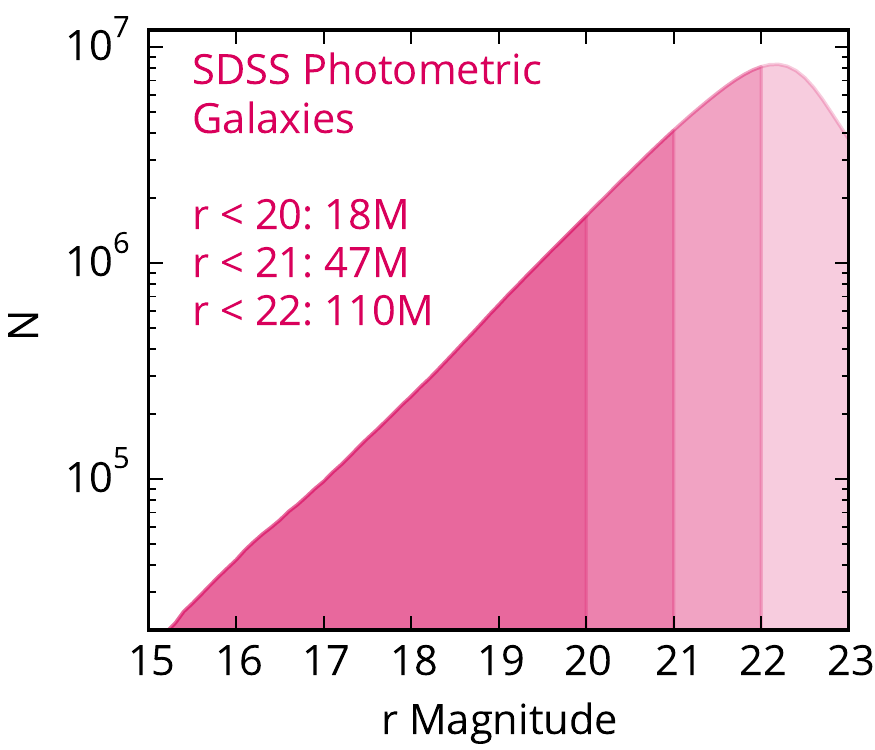}
\caption{
  The r-band magnitude distribution of the SDSS photometric galaxies. In this work, we use three magnitude limits: $r < 20, 21,$ and $22$.  \label{fig:photomagdist}
}
\end{figure}

The SDSS photometric dataset consists of more than 100 million extended sources observed in the $ugriz$ bands over a quarter of the sky \citep{york00,aba09}. The magnitude distribution of these objects is presented in Figure \ref{fig:photomagdist}. Throughout this work, we use the SDSS model magnitudes without any dust reddening correction applied. We will focus on three magnitude-limited samples, for which the total number of sources over the entire SDSS footprint is given by
\begin{eqnarray}
{\rm N}(r < 20) \simeq&1.8\times10^{7} \nonumber\\
{\rm N}(r < 21) \simeq&4.7\times10^{7} \nonumber\\
{\rm N}(r < 22) \simeq&11.0\times10^{7}\;.
\label{eq:sample_size}
\end{eqnarray}
A study of potential systematic effects affecting these samples is presented in \citet{scranton02}. To ensure sufficient overlap between the spectroscopic reference sample and the photometric unknown sample and simplify the analysis, for example by  minimising extinction effects at low galactic latitudes, we focus on a 4800 square degree area within the Northern Galactic Cap, defined by:
\begin{eqnarray}
131\degr < &\alpha& < 241\degr\nonumber\\
5\degr < &\delta& <  60\degr
\label{eq:footprint}
\end{eqnarray}
and discarding small regions with poor photometry surrounding bright stars. This selection leads to samples with roughly one third of the sizes presented in Eq.~\ref{eq:sample_size}. However, the adopted sky coverage is sufficiently large to be representative of the entire SDSS footprint, so the measured properties presented throughout this analysis are applicable to the full photometric sample, with minimal uncertainty due to sample variance. 

Spectroscopic redshift measurements are available for only a few percent of the photometric sources and mostly limited to the brighter objects. Various photometric redshift estimates for these sources have been performed over the years \citep{csabai07,cunha09,sheldon12,brescia14,greisel15}. In this work we will use the photometric redshift catalogue provided by \cite{csabai07}. These redshift estimates are obtained by calibrating the mapping between observed colours and redshift using galaxies for which spectroscopic measurements are available. The mapping is then applied to the entire photometric dataset by sampling it in colour space with a KD-Tree nearest neighbour technique. We use the photometric redshifts estimates given in the SDSS Data Release 10. They were recently updated with an expanded training set in Data Release 12. For the purposes of this work, the differences between the two tabulations are minimal. These photometric redshifts will be used in two ways; in Section~\ref{sect:photsep} we will compare this photometric redshift estimate to our inferred clustering redshifts. In Section~\ref{sect:totdist} we will treat these photometric redshifts as an ``observable'' property of galaxies that we use to sample the entire data set in a manner that optimises the accuracy of clustering-redshift estimation.

\subsection{Clustering-based Redshift Measurement}
\label{subsect:clustz}

We use the procedure presented in \citet{menard13}, tested against simulations by \citet{schmidt13}, against SDSS spectroscopic data in \citet{rahman15a} and applied to 2MASS photometric survey in \citet{rahman15b}. The general approach is to (i) sample the photometric space of a given dataset in small cells, (ii) locate the corresponding objects in redshift space using spatial clustering, (iii) normalise their redshift distribution according to the number of objects in each cell and (iv) combine these individual estimates into a global redshift distribution. The details of the algorithm and implementation is described in these papers, and we refer the reader to them for a detailed discussion. In this section, we briefly re-introduce the main concepts.
We first consider two populations of extragalactic objects:
\begin{enumerate}
	\item We treat the SDSS photometric galaxies as our \emph{unknown} population. 
	This population is characterised by a redshift distribution ${\rm d N_u/d}z$, number of sources ${\textrm N}_u$. Additionally, this population is characterised by a clustering amplitude with respect to the overall matter distribution, or \emph{bias} $\overline{b_u}(z)$. We use the bar to remind the reader that we are considering a bias averaged over a broad range of spatial scales. For this population, the goal is to estimate ${\rm d N}_u/{\rm d}z$. 
    \item We make use of a \emph{reference} population of sources with known angular positions and redshifts. This population is characterised by a redshift distribution $\rm d{\textrm N}_r/\rm dz$, number of sources ${\textrm N}_r$, and a redshift dependent galaxy bias $\overline{b_r}(z)$. We point out that the reference population does not need to be representative of the unknown sources in terms of object type or luminosity but only trace similar large-scale structure patterns.
\end{enumerate}
If the two populations do not overlap in redshift, their angular cross-correlation is expected to be zero (ignoring gravitational lensing effects). In the ideal case of an isolated unknown sample with a very narrow redshift distribution, its redshift location can be recovered by splitting the reference population into contiguous redshift slices $\delta z_i$ and measuring the angular or spatial cross-correlation with the unknown population $w_{ur}(\theta, z_i)$ for each subsample $i$. The redshift distribution of the unknown sample can then be simply normalised by
\begin{equation}
\int \frac{ {\textrm dN_u} }{{\rm d} z}\,{\rm d}z = {\textrm N}_u\;. 
\label{eq:normalization}
\end{equation}
This normalization removes the need to characterise the amplitude of the unknown sample's bias $\overline{b_u}$. In a more general case, when the unknown sample spans a finite redshift range, the redshift evolution of its bias can affect the redshift distribution estimate. But as motivated by \citet{menard13} and demonstrated by \citet{rahman15a}, these uncertainties can be made sufficiently small for many astrophysical applications. First, we note that only the relative redshift evolution of the unknown sample's bias (${\rm d \log\overline{b_{u}}}/{\rm d \log }\, z$), has an effect on the inferred redshift distribution. Additionally, the uncertainty due to the unknown redshift evolution of the bias can be minimised by subdividing the unknown population in the photometric space into subsamples that have narrower redshift distributions.

Similar to the procedure presented in \citet{rahman15a,rahman15b}, we measure the set of spatial cross-correlations over a range of projected radii $0.3 < r_{p} < 3$ Mpc weighted by $r_{p}^{-0.8}$. To avoid measurement errors over large angular apertures on sky and to limit the cosmic variance errors, we restrict our analysis to redshifts greater than $z = 0.03$. To account for on-sky source density variances, we estimate the mean source density ``locally" by measuring it in 16 equal area regions spanning our analysis footprint. We describe the reference spectroscopic sample and its corresponding bias in Appendix \ref{sect:refsamp}. 

While this technique samples a given population in its photometric space for better accuracy, but it is important to note that redshift information is only extracted from the angular position of the sources. It does not rely on any assumption of the SED of the sources, it can be applied at any wavelength, and it does not required any knowledge of the response function of the instrument (such as filter curves and atmospheric transmission). It provides us with a different avenue to characterise the mapping between photometry and redshift, which is the key quantity needed to estimate the redshift of photometric sources.

\begin{figure}[h]
\begin{center}
\includegraphics[scale=1]{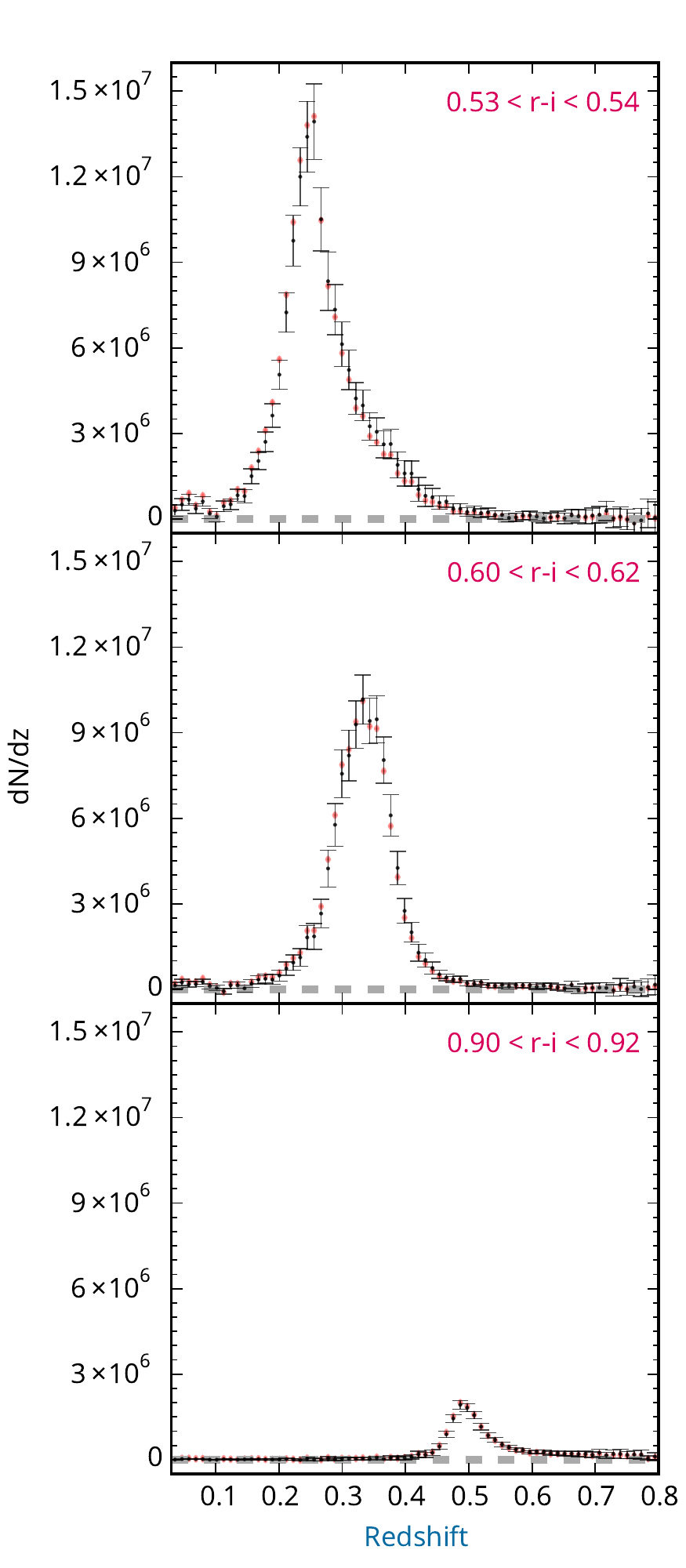}
\caption{
	Example clustering redshift distributions of r-i selected samples with r $<$ 21. 
	Each selected sample leads to a redshift distribution sufficiently narrow ($\sigma_z\sim0.05$) to neglect the possible evolution of the unknown sample's clustering bias. The red diamonds indicate the inferred distribution when correcting for the unknown sample bias from Equation \ref{eq:bias}.
} 
\label{fig:crdplot}	
\end{center}
\end{figure}

\subsection{Examples of Redshift Distribution Estimates}
\label{subsect:exredest}

As an illustration of the method, we first apply the clustering redshift technique to three colour-selected samples:
\begin{eqnarray}
0.53 < r-i < 0.54,~~&\mathrm{with~N}\simeq1.6 \times 10^6\nonumber\\
0.60 < r-i < 0.62,~~&\mathrm{with~N}\simeq1.2 \times 10^6\nonumber\\
0.90 < r-i < 0.92,~~&\mathrm{with~N}\simeq0.2 \times 10^6
\end{eqnarray}
We present the corresponding redshift distribution estimates in Figure~\ref{fig:crdplot}. As can be seen, each subsample can be localised in redshift space. The absence of detected correlation puts strong constraints on the absence of sources at the corresponding redshift. In addition, this example illustrates how the clustering redshift technique can be used to characterise the mapping between source properties, in this case $r-i$ colour, and redshift. It also shows the typical noise level of the measurement as a function of sample size.

We compare the inferred redshift distribution, applying a simple estimate for the redshift dependence of the unknown bias taken to be proportional to $z$ above $z=0.1$ (see Appendix \ref{sect:refsamp} for additional details). As pointed out in Section~\ref{subsect:clustz}, our estimator is only sensitive to the logarithmic redshift derivative of $\bar b_r(z)$ and $\bar b_r(z)$ over the redshift interval of each subsample. Interestingly, we can observe that each selected subsample appears to be distributed over a relatively narrow redshift range, with $\delta z\sim0.05$. Over such a  redshift interval, the redshift evolution of the galaxy bias is expected to be small enough to have no significant effect on the estimated redshift distributions. As quantified in \citet{menard13}, in this regime we expect an uncertainty on the mean redshift of each subsample to be smaller than 0.01. This will be verified in Section~\ref{sec:sampling}. 

\begin{figure*}[t]
\center
\includegraphics[scale=1.0]{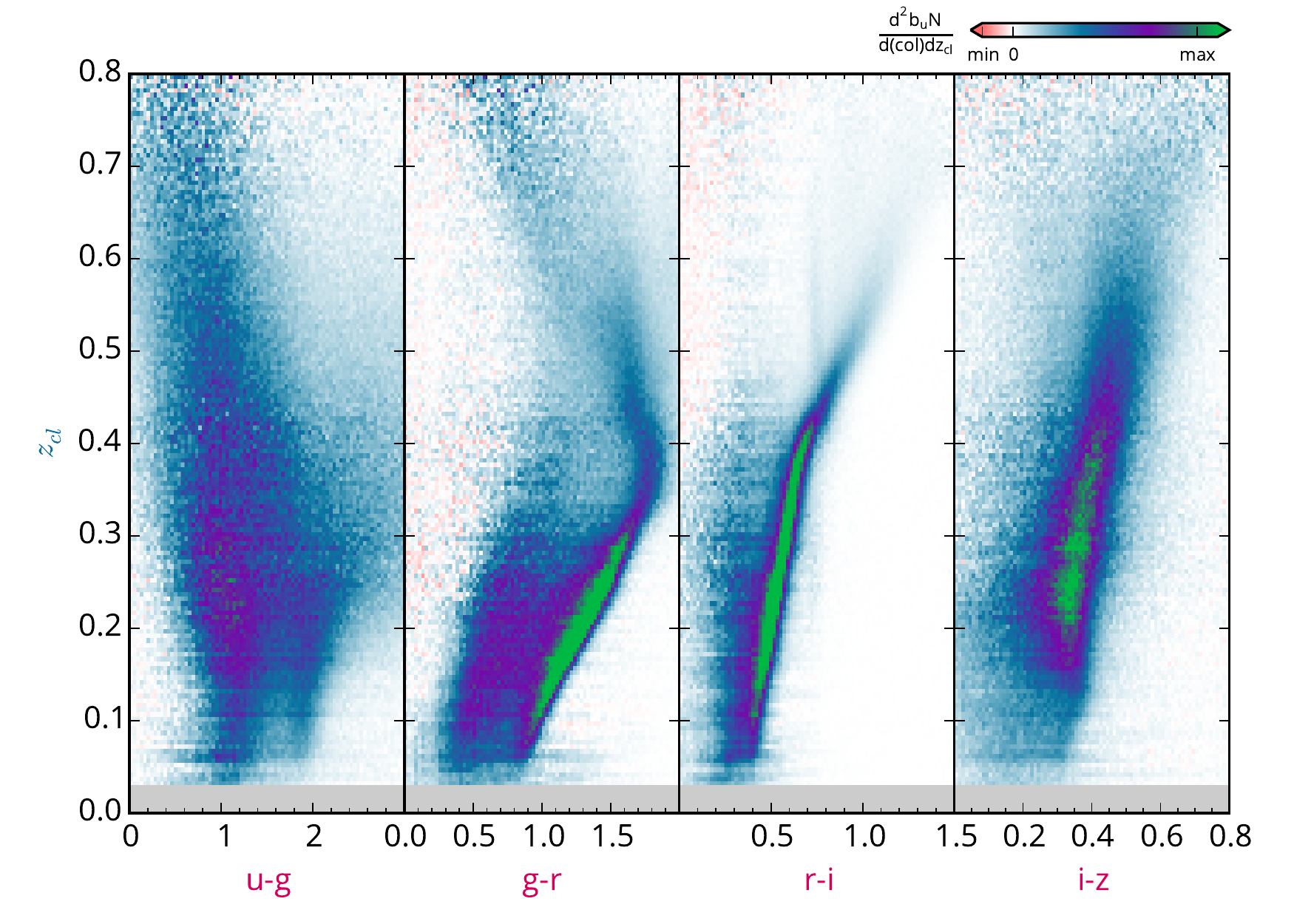}
\caption{The clustering redshift distributions of SDSS extended sources with $r < 21$ as a function of $u-g$, $g-r$, $r-i$, and $i-z$ colours. Each column represents an independent estimate of the redshift-dependent clustering for a sources selected in a narrow color bin. These color-redshift mappings reveal the existence of sources up to $z\sim0.8$, the separation between red and blue galaxies, and the effects of SED features redshifting through the filters, such as the $4000\,$\AA\ break as it passes through the $g$- and $r$-filters at $z = 0.37$.
We note regions where no sources exist, corresponding to forbidden colours for celestial sources at certain redshifts.}
\label{fig:colheatmap}
\end{figure*}

\section{Mapping the Colour-Redshift Relationship\\ with Clustering Redshifts}
\label{sect:colsep}

We now consider extended sources with $r<21$ and characterise the \emph{mapping} between {colours} and {redshift}. We note that once this global mapping has been characterised, it can be applied to individual galaxies to obtain their respective redshift probability distribution functions.

To sample the entire colour space, we first consider each colour separately and split its range into 80 bins. This corresponds to a sampling of roughly 0.02 mag in colour, which is similar to the typical error in the colour measurement. For each of the 80 colour-selected samples, we measure the set of angular cross-correlations with the reference sample, sliced into redshift bins with $\delta z=0.005$ between $z=0.03$ and $0.8$, corresponding to 154 bins.

At this initial step of our exploration, we are not focusing directly on redshift distribution estimates. Instead we first present the cross-correlation amplitude as a function of colour normalised only by the bias $\overline{\textrm b_r(z)}$ of the reference population and the number of galaxies in each sample. In other words, we do not attempt to correct for the redshift evolution of the unknown population's bias at this stage. The corresponding quantity, ${\rm d^2 (\overline{b_{u}}\,N_u)}/{\rm d}(col){\rm d}z_r$, is shown in Figure~\ref{fig:colheatmap}. Each pixel of these maps corresponds to an independent spatial cross-correlation measurement between the selected unknown and reference subsamples over the entire footprint (Eq.~\ref{eq:footprint}). The noise level of this statistical measurement, mostly due to Poisson noise, can be seen in regions with low or no signal; this is exhibited as positive and negative values oscillating around zero in regions sufficiently far from the main overdensities. We also point out that the dynamical range of the four panels is set by their corresponding minimum and maximum values and is therefore different from one panel to another. This range is wider for colours with more concentrated distributions and it limits the ability to display low-level features. For example, the lack of high redshift galaxies in the $r-i$ panel is only apparent and due to the fact that this colour distribution has the highest concentration among the four panels.

The density distributions presented in the figure reveals the colour-redshift relation for the SDSS photometric galaxies. First, we observe that sources with $r<21$ can be found up to at least $z=0.8$, with the highest concentration at $z\sim0.25$. The figure reveals a number regions devoid of galaxies or, in other words, ``forbidden colours'' for celestial objects at certain redshifts. The density distributions appear to follow well-defined sequences as a function of redshift. These sequences are narrower in the $g-r$, $r-i$ and $i-z$ colours. Bimodality in the colour distribution can be seen at $z<0.4$ for the $u-z$, $g-r$ and $r-i$ colours. In the $g-r$ panel, we identify the separation between red (quiescent) and blue (star-forming) galaxies. Additionally, the 4000\ \AA\ break is redshifted from the $g$-band into the $r$-band filter at $z \sim 0.35$ and leads to a prominent turn in the redshift tracks of the related colours, $g-r$ and $r-i$. At $z\sim0.7$ the $4000\,$\AA\ break is redshifted to the $i$ band and, similarly, a break can be seen in the $i-z$ panel at that redshift.

Since this sample is magnitude-limited in the $r$-band, flux measurements at shorter wavelengths for red objects will be more affected by noise. Such effects are apparent when following the red galaxy track in the $u-g$ colour panel. Starting from low redshift, one can observe an excess of clustering for sources redder than the reddest template used here. This is due to noisy photometric measurements. As can be seen, the level of scatter increases towards higher redshifts. At $z\sim0.4$, when the $4000\,$\AA\ break leaves the $g$-band, the $u$- and $g$-magnitude measurements for faint red galaxies become noise dominated and the corresponding track is no longer visible as the clustering signal is diluted over a wide colour range. In contrast, the redshift-colour sequence is detected with sufficient signal-to-noise ratio for star-forming galaxies at those redshifts.

We note that the narrow, horizontal features visible at low redshift appear to be numerical artifacts introduced by our current pixelised density estimate on the sphere. Similarly, the weak vertical overdensity seen at $r-i\simeq0.7$ appears to be due to a slight offset in the estimation of the mean density of the photometric sample for a small range of colour. These will be addressed in the next version of the algorithm (Mendez et al. in prep). For the present analysis, these features are sufficiently weak and will not significantly affect our final redshift distribution estimates.

\begin{figure*}
\center
\includegraphics[scale=1.0]{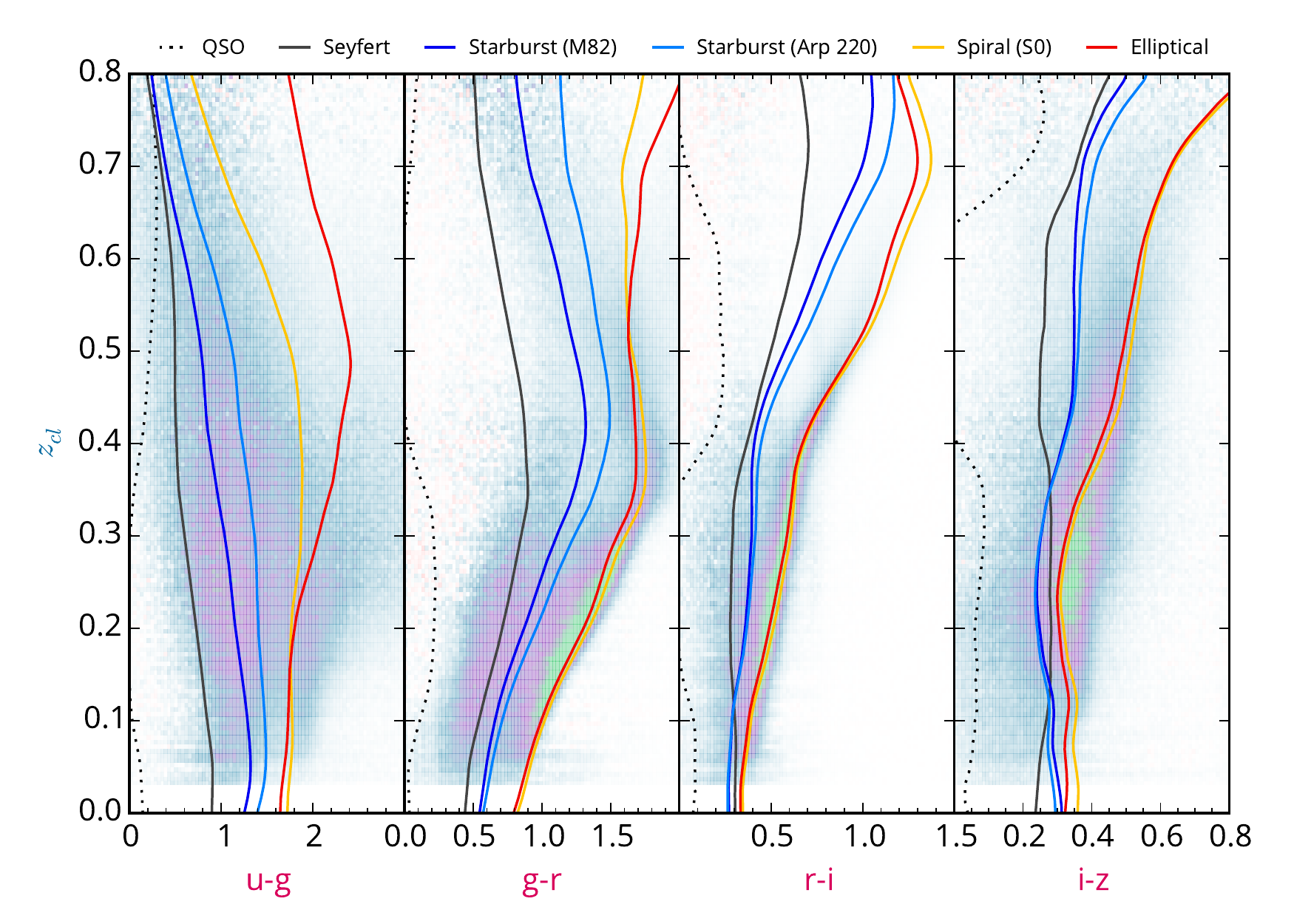}
\caption{The colour-redshift mapping (as in Figure \ref{fig:colheatmap}) with redshifted galaxy templates from \citet{polletta07} convolved with the SDSS passbands. The blue tracks are the colours inferred for starburst galaxies, specifically Arp 220 and M82, and the red tracks show synthetic spiral (S0) and elliptical (13 Gyr) galaxy templates. The black tracks are from a composite Seyfert 1.8 (solid) and Type-1 QSO (dotted) templates.
 \label{fig:colheatmap_tracks}}
 \end{figure*}

The density distributions presented in the figure visually indicate the amount of redshift information that can be extracted from a given colour. The strongest correlations between these two quantities are found in the $r-i$ and $g-r$ colours, as expected, since the strongest large-scale feature in galaxy SEDs at these wavelengths is the $4000\,$\AA\ break. We note that this colour-redshift mapping can be used to infer the redshift probability distribution of a given galaxy, based on its measured colours. 

We note that the figure only shows density distributions for four one-dimensional projections of the global colour distribution. Degeneracies exist in the relationships between redshift and individual colours but some of them can be broken by simply combining information from multiple colours. For example, in the $g-r$ panel, sources with colours around $g-r\sim1.2$ map onto bimodal redshift distributions with $z\simeq0.2$ and $z\simeq0.5$ but adding information from the $r-i$ and/or $i-z$ colours can be used to differentiate each redshift region. A more optimal  estimation can be made by taking into account all the colour information requires sampling the full four-dimensional distribution. In addition, information from other photometric dimensions (such as brightness, size, light profile, ellipticity, or environment) can be added to further refine the mapping between photometric properties and redshift. This will be considered in future studies.

\begin{figure*}[t]
\center
\includegraphics[scale=1.0]{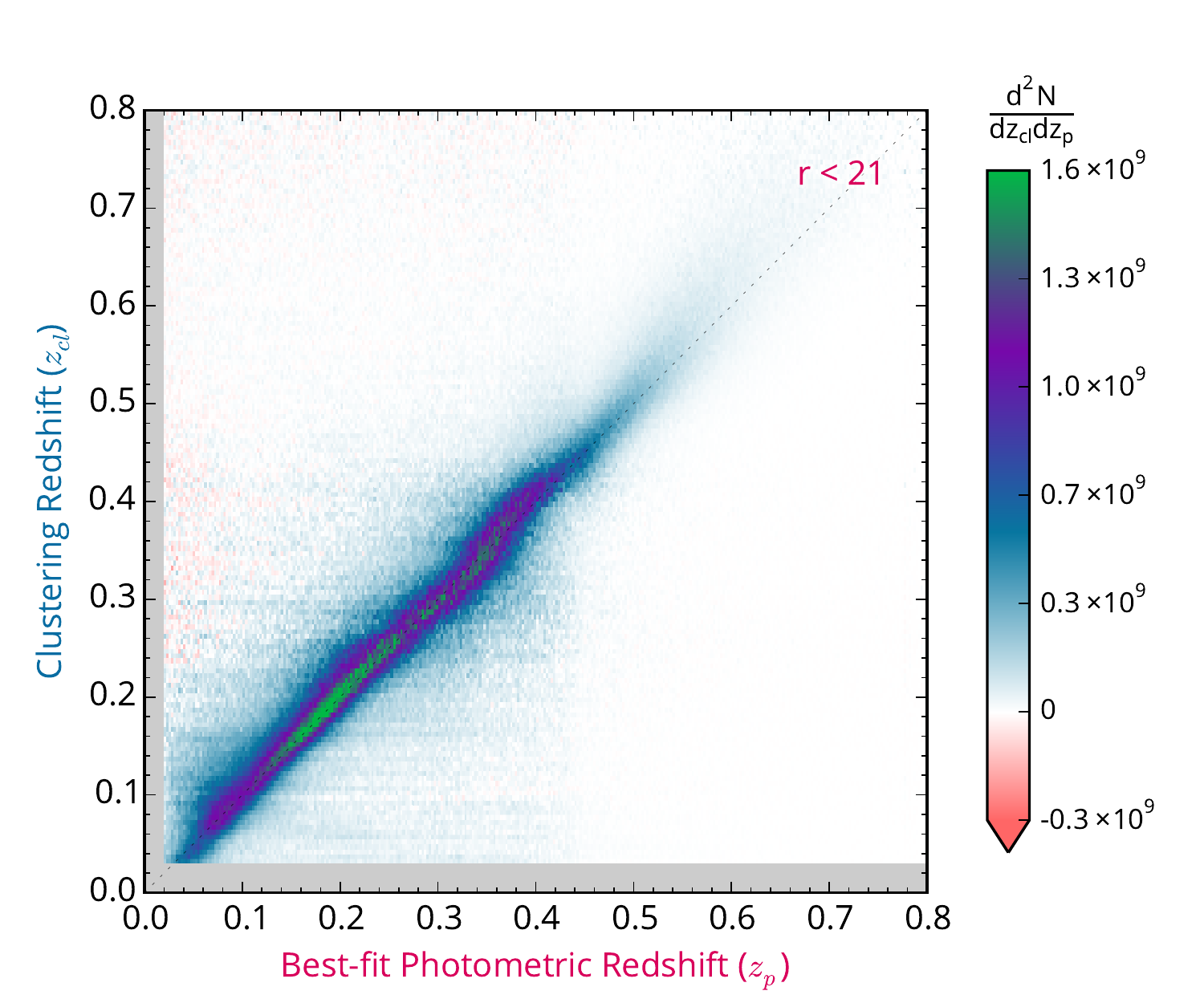}
\caption{The clustering redshift distribution of $r<21$ extended sources as a function of their best-fit photometric redshift from the KD-tree method \citep{csabai07}. The one-to-one relationship is indicated by the grey dotted line.
} 
\label{fig:fullheatmap}
\end{figure*}

\subsection{Comparison to Spectral Energy Distribution Templates}

To obtain more insight into the colour-redshift mapping characterised through clustering measurements, we compare our results to colour-redshift tracks obtained by convolving \emph{non-evolving} SEDs with the SDSS filter curves. As an illustration we use a set of galaxy and Active Galaxy Nuclei (AGN) templates from \citet{polletta07} to highlight the colours of different source types. We show the corresponding tracks in Figure~\ref{fig:colheatmap_tracks}. We also considered templates from the GALEV series of simulations \citep{kotulla09} and the population synthesis models of \citet{bruzual03} but as their global behaviours appear to be similar, we do not show them for simplicity. The comparison between the redshifted templates and the clustering redshift results provides population-level information about the relative prevalence of specific templates at any given redshift.

\textbf{Similarities between the two techniques:~} Overall, we observe that the tracks obtained from the SED templates display trends similar to the density distribution inferred through clustering measurements: the span in colours, the orientation of the tracks, the main breaks, and forbidden regions are in good agreement. This shows that clustering-based redshift inference is able to recover the overall shape of the sources' SEDs. For example, around $z=0.4$, the red tracks are in good agreement with the measured density distributions, indicating that the shape of the SEDs around the 4000~\AA\ break is reproduced with high fidelity. The computed curves can be used to highlight the relative contributions of ``red'' and ``blue'' galaxies as a function of redshifts. In particular, they reveal the existence of degeneracies in the colour-redshift mapping due to galaxy type. A given colour can map onto two galaxy types at different redshifts; for example, the apparent $g-r$ colour of star-forming galaxies at $z\sim 0.3$ is the same as that of quiescent galaxies at $z \sim 0.05$. As mentioned above, this degeneracy can be broken by adding information from the $u-g$ colour, or alternatively morphological type obtained from the galaxy shape or light profile.

As mentioned earlier, the $u$-band observations are the least sensitive, which propogates into the $u-g$ colours as an increased uncertainty. We note that the quiescent galaxy tracks in this colour are insufficiently red for certain galaxies in this space, which is likely due to the photometric uncertainty rather than an issue with the template SEDs. 

The redshift-colour relationships contain information about the relative distribution of star-forming and quiescent galaxies at a given redshift. In regimes where all sources are detected in all bands, we can extract information about the ratio between the two types of galaxies; for instance, at $z\sim0.2$, the populations can be separated into red and blue galaxies. This property is instructive for the development of photometric redshifts, which require information about the relative contributions of the galaxy type over a full population.

\textbf{The contribution of M-stars:~} In the $i-z$ panel, all the computed tracks are found to lie on the left side of the measured density distribution in the range $0.15 < z < 0.4$. As this effect is not seen in the $r-i$ panel, this suggests that it is due to a feature present in the $z$ band. This points to the existence of a feature responsible for a deficit of flux at a rest wavelength around $7000\,$\AA. This turns out to match the strong TiO absorption band at 7100\ \AA. This absorption bands is a broad, prominent feature that is especially visible in the spectra of galaxies and arising from the flux contribution of late, M-type stars. The TiO band disappears in the spectra of K-type stars; consequently, a slight variation in the ratio of K-type to M-type stars will significantly affect the prominence of this absorption band. Our results suggest that the set of SEDs presented here is not representative of the majority of galaxies and that the typical fraction of M-stars is lower at those redshifts. This deviation is not expected to have a strong effect on photometric redshift inference as this feature is weaker compared to the prominent 4000~\AA\ break but it is likely to introduce a small bias.

\textbf{Redshift dependence of SEDs Templates:~} If the redshift evolution of an SED is not properly accounted for, we expect an overall tilt between the computed tracks and the measured signal. Figure~\ref{fig:colheatmap_tracks} indicates that such an effect is apparent for the two quiescent templates in the $g-r$ color. As this effect is not seen in the $r-i$ color, it suggests that the chosen \emph{static} SEDs require a redshift dependent shape blueward of the $4000\,$\AA\ break.

\textbf{Active Galactic Nuclei:~} In additional to using normal galaxy templates, we also present the redshift-colour tracks for AGNs. These sources are typically much bluer in all colours than tracks for normal galaxies. Interestingly, these tracks seem to follow the blue envelope of the distributions at $z<0.4$ where a large fraction of star forming galaxies are detected. This shows that AGN activity is can lead to a range of flux contributing to the overall flux of a galaxy. We have verified the identity of a small subsample of galaxies in this portion of the colour-redshift space through a comparison with the spectroscopic subsample within SDSS; these galaxies are consistently spectroscopically typed as AGN hosts with some highly star-forming sources. This analysis demonstrates the necessity of a diversity of templates to fully reconstruct the colour-redshift behaviour of SDSS galaxies including active galaxies. At $z>0.5$ one can observe a lack of clustering-redshift signal along the QSO tracks shown as dotted lines. This is in line with the fact that the photometric sample only contains extended objects; it is not expected to include a substantial fraction of QSOs at higher redshifts, where the host galaxy is not detectable. We note that the wiggles appearing in the $r-i$ and $i-z$ tracks at $0.35< z < 0.7$ are due to presence of the H$\beta$ emission line in the $i$-band filter.

\textbf{Dust:~} We expect the effects of dust extinction to redden the apparent colours of sources with respect to the SED-based tracks. A sheet of dust, for example distributed within the Milky Way, would produce a systematic rightward shift in all colours. As no overall offset is seen between the measured density distributions and the computed tracks, we do not find any indication for the need of significant foreground dust correction as expected with the choice of the footprint defined in Eq.~\ref{eq:footprint} and located at high Galactic latitudes. In the $g-r$ and $r-i$ planes, we do not observe any sources at colours redder than those predicted by the tracks computed for quiescent galaxies. This suggests that the (low) level of dust used in the corresponding SED templates is appropriate. The same statement applies to the $i-z$ panel, after correcting for the strength of the TiO band described above. For the blue star-forming galaxies, we do not have the same constraint; the location of these galaxies in colour-redshift space may be affected by dust extinction, but this effect appears degenerate with the diversity of galaxy SEDs. We do observe an excess of reddened galaxies in the $u-g$ colours at low redshift but, as described above, this is caused by photometric noise: red galaxies do not display much flux blueward of the $4000\,$\AA\ break, leading to noise-dominated colour estimates.

\begin{figure*}
\center
\includegraphics[scale=1.0]{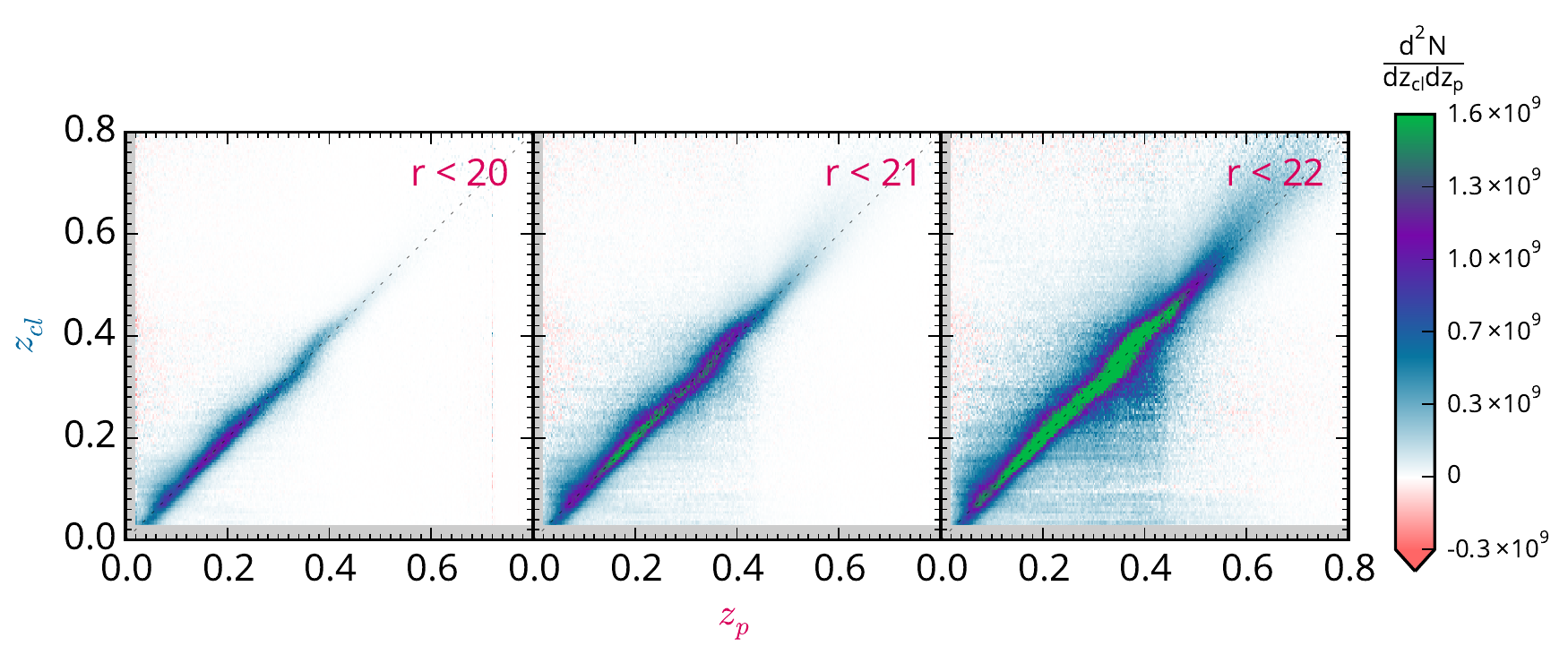}
\caption{
The clustering redshift distribution of extended sources as a function of their best-fit photometric redshift for limiting magnitudes of $r < 20$ (left), $r < 21$ (middle), and $r < 22$ (right). The one-to-one relationship is indicated by the grey dotted line. The peak of the distribution is clipped in the $r <22$ sample to better reveal the broadening of the distribution with increasing limiting magnitude.
} 
\label{fig:threeheatmap}	
\end{figure*}

\section{Comparison to Photometric Redshifts}
\label{sect:photsep}

We now compare our clustering-based redshift estimates to photometric redshifts estimates. As introduced in Section~\ref{subsect:sdss}, in this analysis we make use of the KD-Tree nearest neighbour photometric redshift estimates from \citet{csabai07} which are calibrated using a set of observed galaxy spectroscopic redshifts. To perform a comparison in a simple manner, we sample the photometric sources according to their best-fit photometric redshift $z_p$. We do so by splitting the photometric sample into 320 $\delta z_p = 0.025$ bins, from $z_p=0.02$ to $0.8$. This time, for each subsample, we estimate actual redshift distributions ${\rm {d^2N_u}}/{{\rm d}z_{p}{\rm d}z_{cl}}$ by computing $\overline{{w_{ur}}}(z) / (\overline{b_{u}}\,\overline{b_{r}})$, normalised by the number of sources in each subsample. We use the unknown bias estimate as discussed in Section~\ref{subsect:exredest}. 

Figure~\ref{fig:fullheatmap} shows the estimated clustering-redshift distribution  ${\rm {d^2N_u}}/{{\rm d}z_{p}{\rm d}z_{cl}}$ for the SDSS extended sources with $r<21$ as a function of best-fit photometric redshift. The density distribution is normalised such that its integral over the entire $z_p$ and $z_c$ redshift ranges yields the number of selected photometric sources. Overall, we observe a good agreement between photometric redshift and clustering redshift estimates, with the majority of the signal centred along the diagonal with a width ranging from $\sigma_z\simeq0.03$ to $\sigma_z\simeq 0.05$ at higher redshift. At low photometric redshift values one can observe a higher noise level (indicated by the higher fraction of red pixels), which is due smaller sample sizes. of $\Delta z \sim 0.05$. At $z\sim0.35$, we find a deviation from the one-to-one relation corresponding to the redshift at which the $4000\,$\AA\ break lies between the $g$ and $r$ bands, making photometric redshift estimation more difficult. 
This effect was present as a color-redshift degeneracy in the $g-r$ colour in Figure \ref{fig:colheatmap_tracks}. 
At higher redshift, we note that there is a tendency for the photometric technique to consistently infer redshifts lower than those found with the clustering redshift technique. This is likely due to the training set used to infer the photometric redshifts, which will be dominated by lower redshift sources. We point out that the observed width of the measured redshift distributions is driven by the intrinsic accuracy of the photometric redshift estimation. The intrinsic width of the clustering redshift technique, in other words its response function for sources located at a fixed redshift, is of order $\Delta z_c \sim 10^{-3}$ \citep{rahman15a}, an order of magnitude smaller

Figure~\ref{fig:threeheatmap} shows how the mapping between best fit photometric redshift $z_p$ and clustering redshift $z_c$ as a function of limiting magnitude, for $r<$ 20, 21, and 22. We can observe the mean redshift of the samples to increase with fainter sources: from $\langle z\rangle \sim 0.2$ to $0.3$ for $r < 20$ to $22$. 
As with the previous figure, we note that at high redshift, the photometric redshifts slightly underestimate the redshifts of the sources. We also observe an increase in the width of the distribution and the fraction of off-diagonal signal at fainter limiting magnitude, due to increase in the photometric noise. We note that the clustering redshift technique, extracting information from source positions, is not affected by photometric noise (apart from the way the dataset is being sampled). 

\begin{figure*}[ht]
\center
\includegraphics[scale=1.0]{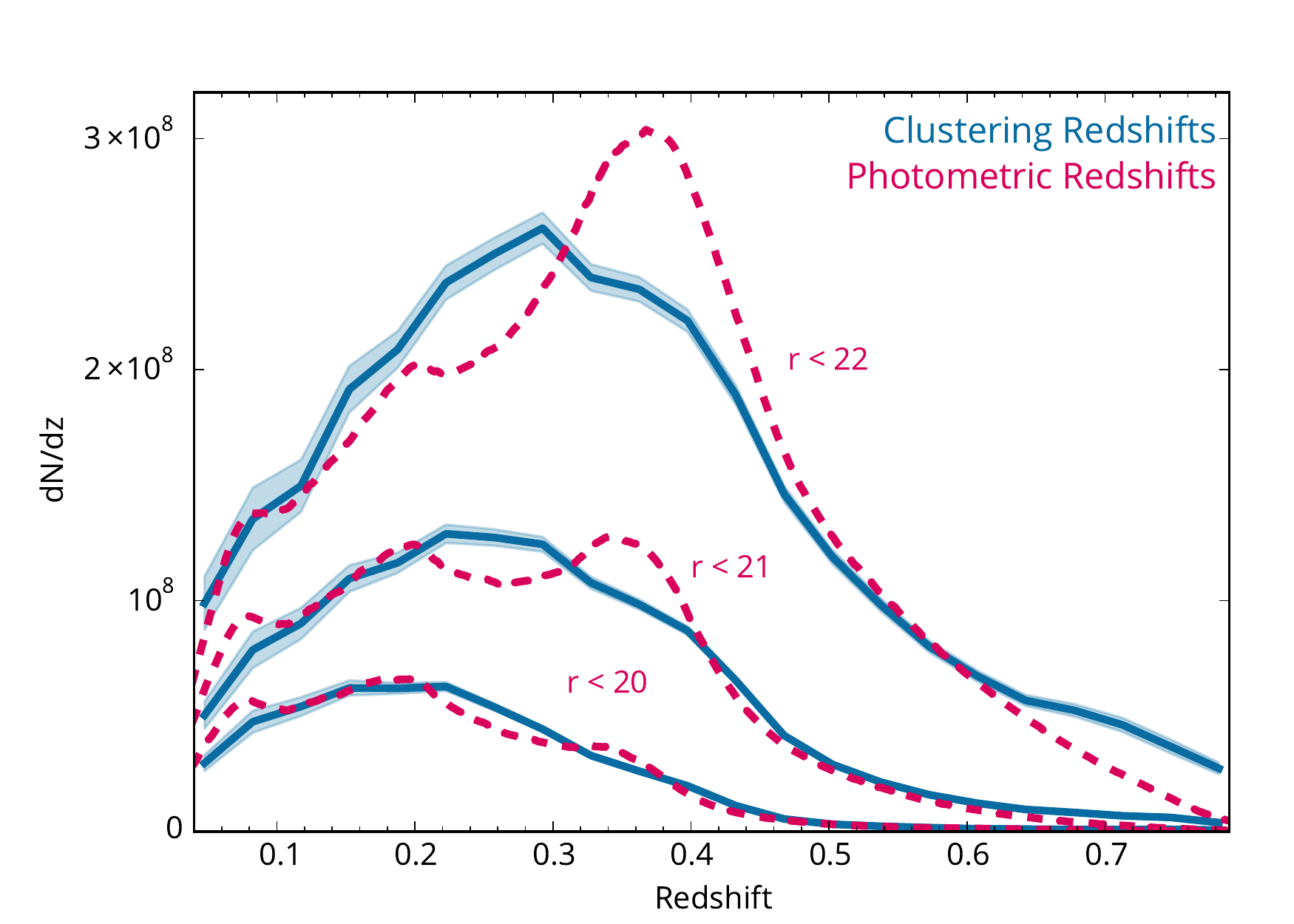}
\caption{
	The global clustering redshift distributions of SDSS extended sources as a function of limiting magnitude (solid blue curves). For comparison, the KD-tree photometric redshift estimates are also shown (dashed magenta curves). The clustering redshift estimation leads to smoother distributions, with no indication of the structure suggested by the photometric redshift estimates (likely artifacts imprinted by their spectroscopic training set). The blue bands indicate the standard deviation of the distributions as estimated by their on-sky variance. 
}
\label{fig:totdist}	 
\end{figure*}

\section{The Global Redshift Distribution\\ of SDSS galaxies}
\label{sect:totdist}

Having estimated clustering-based redshift distributions for the SDSS photometric galaxies subsampled as a function of either their colours (Section~\ref{sect:colsep}) or photometric redshift (Section~\ref{sect:photsep}), we now estimate the global redshift distribution of the parent sample. As discussed in \citet{menard13}, the best strategy to follow is to use a sampling that minimises the redshift range of each subsample. Comparing the results obtained in the two previous section, it is clear that using the best-fit photometric redshift is more optimal than each individual colour-based sample. As described above, this leads to a series of photometric subsamples for which the redshift width is of order $\sigma_z=0.05$. This ensures that the redshift evolution of the unknown bias $\overline{ b_u}(z)$ is negligible over such a range and its functional form has therefore little effect on the corresponding ${{\rm d}{\textrm N}_u}/{\rm d}z$ estimate. Sampling the parent sample as a function of individual colours, as in Section~\ref{sect:colsep}, does not provide samples sufficiently narrow in redshift to neglect the redshift dependence of $\overline{b_u}$. We note that sampling the parent population in its 4-dimensional colour space would actually lead to subsamples for which the redshift distributions are similarly narrow to those obtained through photometric redshift sampling. Our choice of using photometric redshifts to sample the population is only driven by simplicity. Photometric redshift estimates, in general, are \emph{not} required for clustering-redshift estimation. They are only used to provide a straightforward, one-dimensional sampling of the data. We also note that results could be even less sensitive on the evolution of $\overline{b_u}$ by sampling additional dimensions of the photometric space, such as brightness, size, concentration, and ellipticity.

Having chosen $z_p$ as the parameter over which we sample the dataset, we now determine the global redshift distribution by simply summing each of the density distributions displayed in Figure \ref{fig:threeheatmap} along the $z_{p}$ axis:
\begin{equation}
\frac{\rm d N_u}{{\rm d}z_{cl}} = 
\int \frac{ {\rm d^2\,N_u}}{{\rm d}z_{p}{\rm d}z_{cl}} {\rm d}z_{p}\;.
\end{equation}
We present the corresponding redshift distributions for the three magnitude-limited samples $r<20,\;21$, and $22$ in Figure \ref{fig:totdist} as blue lines, with a resolution of $\delta z_{cl} = 0.035$. The uncertainty in the distribution shown with blue regions is estimated through jackknife error estimates using 16 regions to sample the overall footprint. The uncertainty is larger at high redshifts due to small reference sample sizes leading to greater noise in the cross-correlation measurement, and higher at low redshift as well due to the additional effect of cosmic variance. The redshift distributions are found to peak at $z\simeq0.20, 0.25$ and $0.30$ for the three different limiting magnitudes and display tails extending up to at least $z\sim0.8$.

For comparison, we also present the photometric redshift distributions as red dashed lines. To estimate these functions, we only use the value of the best-fit photometric redshift $z_p$ for each source. We do not include their entire probability distribution functions, which are expected to have negligible effects as the typical values of $\sigma_{z_p}$ are of order $0.01$, a scale much smaller than any structure observed in the overall redshift distributions. Globally, we find general agreement for each limiting magnitude between the two redshift estimation techniques, but two differences appear:

\textbf{(i)} The clustering-redshift distributions are significantly smoother than those indicated by the photometric redshift estimates. The latter display some structure with three redshift peaks, at $z\simeq0.07$, $0.20$ and $0.35$ seen at each limiting magnitude. These overdensities are likely caused by redshift attractors originating from redshift overdensities present in the spectroscopic sample used to calibrate the photometric redshifts. This spectroscopic sample is known to display pronounced specific peaks in its redshift distribution: $z\simeq0.07$ corresponds to the Sloan Great Wall (located between $9 \textrm{h} < \alpha < 14 \textrm{h}$ and $-4\degr < \delta < 1\degr$, \citealt{gott05, pimbblet11}) and $z\simeq0.35$ corresponds to the highest density of luminous red galaxies from the SDSS legacy spectroscopic sample, resulting from choices in the spectroscopic target selection. Acting as redshift attractors, these features artificially enhance the density of objects with photometric redshifts matching those values. To conserve the total number of objects in the distribution, these attractors create deficits of sources around the peaks. The presence of these deficits may cause the appearance of a photometric redshift peak at $z\sim0.2$. This effect has been previously discussed in \citet{cunha09}. 
We note that, given the large area considered, such large-scale matter overdensities (of order 10\% over scales of hundreds of Mpc) are not expected to be present. This is in agreement with the clustering redshift estimate which shows no significant redshift substructure for the three different samples. We note that even if overdensities exist in the spectroscopic reference sample at certain redshifts, they are not expected to leave a direct imprint on the clustering redshift estimation as the clustering signal does not depend on the \emph{number} of spectroscopic reference sources but on their angular clustering properties.

\textbf{(ii)} Another difference between the two redshift distribution estimators is seen at high redshift: the clustering redshift distributions appear to have more prominent tails than those inferred by the photometric redshift technique. This is particularly evident with the $r < 22$ sample, where the photo-$z$s predict a negligible number of galaxies at $z \sim 0.8$, while clustering redshifts indicate the existence of sources beyond $z = 0.8$, the maximum redshift considered in this analysis. This effect is due to the skewing of the redshift distributions to lower values, as mentioned in the previous section, and may be caused by the imprint of the redshift distribution of the spectroscopic training set, lacking high redshift sources.

\begin{figure*}
\center
\includegraphics[scale=1.0]{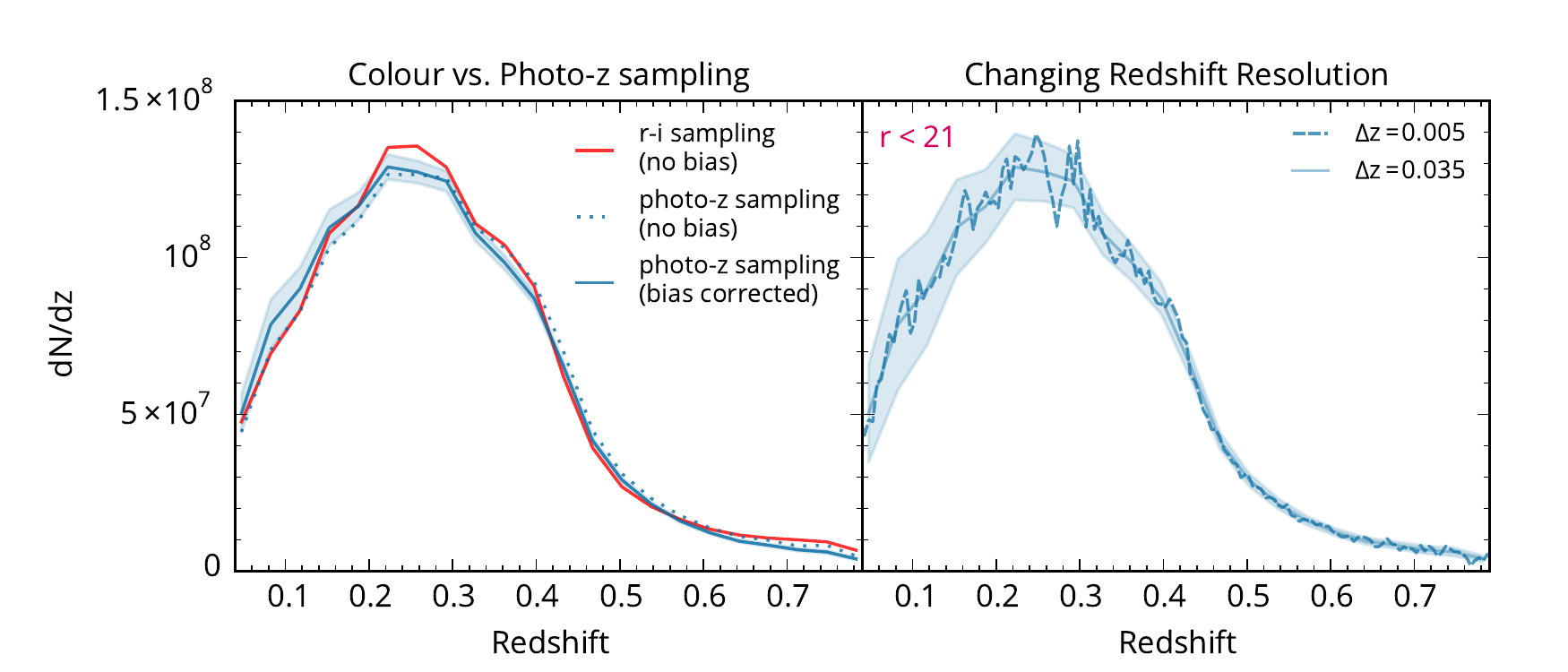}
\caption{
	 \textbf{Left:} 
	 Different estimates of clustering redshift distributions for the $r<21$ sample: with bias correction applied (solid blue), without bias correction (dotted blue), and using the $r-i$ colour to sample the data as opposed to best-fit photometric redshift (solid red). The consistency between the three curves demonstrate that the bias correction has little effect on the final redshift distribution estimates and that direct sampling in colour space can lead to comparable results.
	 \textbf{Right:} Estimate of the clustering redshift distribution with a redshift resolution increased from $\Delta z = 0.035$ (solid line) to $\Delta z = 0.005$ (dashed line). The shaded region shows the uncertainty corresponding at the higher resolution and indicating that the small-scale fluctuations are consistent with noise.
} 
\label{fig:totdist_comparecol}	
\end{figure*}

\subsection{Sampling Considerations\\ with Clustering redshift Estimation}
\label{sec:sampling}

As mentioned above, we have used photometric redshift estimates to sample the photometric data with unimodal and narrow redshift distributions. This approach minimises the sensitivity of the estimator on the redshift evolution of the reference and unknown biases which are expected to be negligible in this case. To demonstrate this, we show the estimated clustering redshift distribution with and without the bias correction for the unknown population. The results are shown in the left panel of Figure~\ref{fig:totdist_comparecol} and demonstrate that the redshift evolution of the bias of the unknown population has virtually no effect on the final redshift distribution estimate.

In addition, we show the typical results that can be obtained by sampling the data directly in colour space. Figure~\ref{fig:colheatmap} shows that the $r-i$ panel displays the strongest correlation between colour and redshift. We then use this colour to sample the data and show the corresponding clustering redshift distribution as the red line in the left panel of Figure~\ref{fig:totdist_comparecol}. Here, we have not applied any normalization for the unknown bias redshift evolution. Interestingly, the red curve displays a rather good agreement with the two blue curves. The estimate using the $r-i$ sampling leads to a slight overestimation of the amplitude of the peak of the distribution at $z = 0.25$ and a deficit a low redshift. This difference is likely caused by the wide redshift distributions of the samples with $r-i < 0.4$ (as seen in Figure \ref{fig:colheatmap}), which will overweight the contribution of the higher-redshift sources without the appropriate bias correction. This problem is even more exaggerated in the $u-g$ and $i-z$ colours, where most samples contain a broad range of redshifts. However, this approach shows that similar results can be obtained by sampling the dataset in a variety of manners. In addition, it further demonstrates that photometric redshift estimates are not required to use the clustering redshift technique.

Finally, as the intrinsic width of the clustering redshift estimator is of order $\delta z=0.001$ \citep{rahman15a}, we point out that this technique has the ability to detect potential narrow peaks in the redshift distribution of a sample, specifically on scales smaller than those on which the unknown population bias is expected to vary. In order to probe the potential existence of such peaks, or small-scale redshift structure, we estimate the clustering redshift distribution using even smaller redshift bins: $\delta z = 0.005$, which is seven times smaller than those presented above. The corresponding results are shown in the left panel of Figure \ref{fig:totdist_comparecol}. Similarly, we estimate the noise level by creating 16 jackknife samples over the footprint and show the corresponding standard deviation with the blue region. As can be seen, no significant small-scale structure is detected and the observed fluctuations are consistent with noise and sample variance due to the intrinsic structure of the spectroscopic reference sample.

\section{Conclusion}
\label{sect:conclusion}

The Sloan Digital Sky Survey has produced a sample of over 100 million extended photometric sources with $r<22$ distributed over a quarter of the sky. Using clustering-based redshift inference, we have explored the redshift distribution of these sources as a function of brightness and colour, following the implementation presented in \citet{rahman15a}. This technique allows us to infer redshifts without any assumption of the sources' spectral energy distributions nor the knowledge of the response function of the system (including filter curves and atmospheric transmission). Our main finding are as follows:\\

$\bullet~$ We have mapped the relationships between individual colours and redshift, allowing us to identify  star-forming and quiescent galaxy populations, AGNs, and colour changes due to spectral features, such as the 4000\ \AA\ break, present in certain passbands and redshift intervals.\\

$\bullet~$ Globally we find a good agreement between our colour-redshift mapping and results obtained by convolving different types of galaxy SEDs with a characterization of the SDSS filter curves. Our analysis also reveals interesting differences, such as the need for a lower fraction of M-type stars (giving rise to TiO absorption bands in the $z$-band) in the selected galaxy templates, and the need for AGNs to reproduce all colours as a function of redshift.\\

$\bullet~$ We have compared our results to photometric redshifts estimates provided by the KD-Tree Nearest Neighbour algorithm. Overall, we find general agreement for each limiting magnitude considered. We also observe deviations between the two estimators at $z\sim0.35$, where the $4000\,$\AA\ break lies between two filters, making color-based photometric estimation more difficult, an effect that does not affect clustering-redshift inference. At higher redshift, there is a tendency for the photometric technique to consistently infer redshifts lower than those found with the clustering redshift technique, an effect likely imprinted by the spectroscopic set used by the KD-Tree Nearest Neighbour.
\\

$\bullet~$ Finally, we present the global clustering redshift distribution of all SDSS extended sources, as a function of limiting magnitude. We find distributions with mean redshifts ranging from 0.1 to 0.3 for sources with limiting magnitudes between $r<20$ and $22$, and tails extending to $z\sim0.8$. While the corresponding photometric redshift distributions indicate a number of spurious redshift features, the clustering redshift distributions show smooth curves with no indication of substructure, as expected with our cosmological model. We also showed that the (unknown) redshift dependence of the clustering amplitude, or bias, of the photometric sources does not have a significant impact on the final clustering redshift distribution estimates.
~\\

This work has provided us with a mapping between colours and redshift for which we have considered four 1-D projections of the colour space. It is a first step towards a full characterization of the mapping between photometric properties and redshift. Future implementations of the technique will allow a direct sampling of the data in its full 4D colour space (for which the use of photometric redshifts in this analysis was a proxy) as well as the use of other dimensions of the photometric space (brightness, size, concentration, ellipticity, and environment) and redshift. This will further weaken the dependence on the redshift evolution of the bias of the photometric population and lead to more precise redshift determinations.

While classical SED-based photometric redshifts have played a major role in extragalactic astronomy, and will continue to provide valuable redshift information with upcoming surveys, we have demonstrated that clustering-based redshift inference is potentially as powerful. The limitations and systematic effects of these two techniques are fundamentally different, enabling complementarity between the two. As the number of measured spectroscopic redshifts for galaxies and quasars increase with time, the accuracy of the clustering redshift technique will keep improving. As surveys increase in their depth and source diversity, the ability to infer redshifts without SED knowledge will be critical to exploring these observations, possibly unveiling phenomena yet to be discovered.

\acknowledgments 

We thank Robert Lupton, Alex Szalay, Jim Gunn, Masataka Fukugita, and Ani Thakar for useful discussions throughout this project. This work is supported by NASA grant 12-ADAP12-0270, the National Science Foundation grant AST-1313302 and the Packard Foundation.

Funding for SDSS-III has been provided by the Alfred P. Sloan Foundation, the Participating Institutions, the National Science Foundation, and the U.S. Department of Energy Office of Science. The SDSS-III web site is http://www.sdss3.org/.

SDSS-III is managed by the Astrophysical Research Consortium for the Participating Institutions of the SDSS-III Collaboration including the University of Arizona, the Brazilian Participation Group, Brookhaven National Laboratory, University of Cambridge, Carnegie Mellon University, University of Florida, the French Participation Group, the German Participation Group, Harvard University, the Instituto de Astrofisica de Canarias, the Michigan State/Notre Dame/JINA Participation Group, Johns Hopkins University, Lawrence Berkeley National Laboratory, Max Planck Institute for Astrophysics, Max Planck Institute for Extraterrestrial Physics, New Mexico State University, New York University, Ohio State University, Pennsylvania State University, University of Portsmouth, Princeton University, the Spanish Participation Group, University of Tokyo, University of Utah, Vanderbilt University, University of Virginia, University of Washington, and Yale University. 

Facilities: \facility{Sloan Digital Sky Survey}

\bibliography{rs-sdsspz}

\appendix



\begin{figure*}
\includegraphics[scale=0.95]{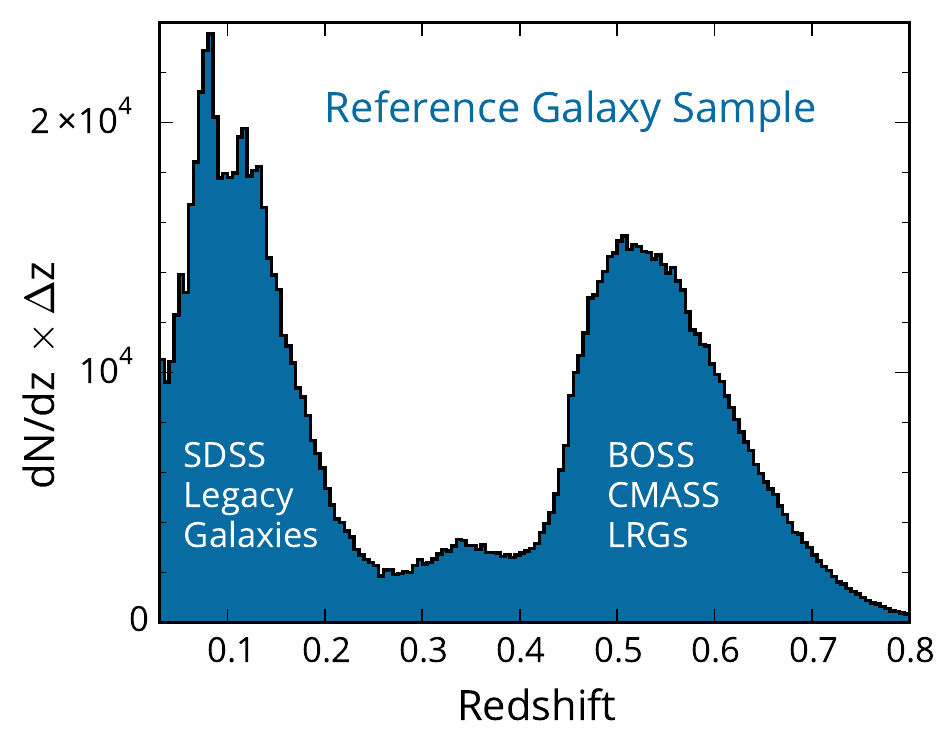}
\includegraphics[scale=0.95]{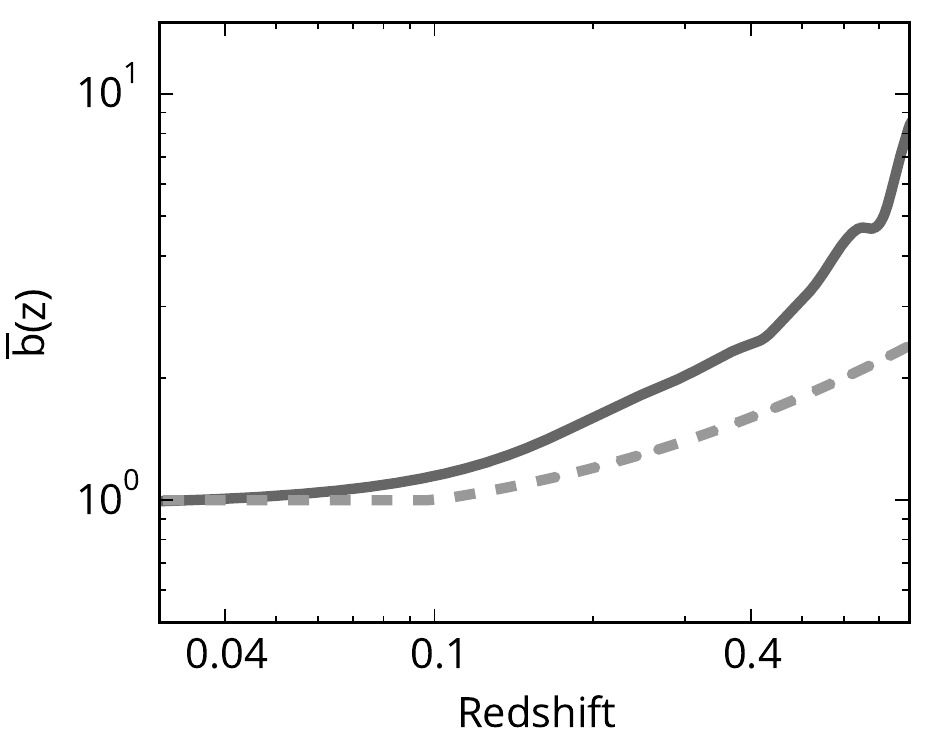}
\caption{
    {\bf Left:} The redshift distribution of the spectroscopic reference population. The population is composed of two samples to span the redshift range: the SDSS Legacy Galaxies at low redshift ($z \lesssim 0.45$) and the BOSS CMASS LRGs at high redshift ($z \gtrsim 0.45$).
	{\bf Right:} {\it Solid line}: The luminosity-weighted galaxy bias $\overline{b_{r}}(z)$ of the spectroscopic reference sample as determined from the \citep{zehavi11} fitting function. {\it Dashed line}: The galaxy bias we use for the unknown samples in this work from Equation \ref{eq:bias} ($\overline{b_{u}}(z)$). 
	As demonstrated in Fig.~\ref{fig:totdist_comparecol}, the final clustering redshift estimates do not strongly depend on functional form used for this quantity
\label{fig:reffig}	
} 
\end{figure*}

\section{Reference Sample Distribution and Galaxy Biases}
\label{sect:refsamp}

For this work, we use spectroscopy from the SDSS \citep{ahn13} survey as a reference sample, specifically combining the SDSS Legacy spectroscopic sample extending up to $z \simeq 0.45$ \citep{strauss02,eisenstein01}, and the CMASS luminous red galaxies from the Baryon Oscillation Spectroscopic Survey extended up to $z \simeq$ 0.8 \citep{padman12}. We present the redshift distribution of this sample on the left of Figure \ref{fig:reffig}. 

Additionally, the clustering-based redshift technique requires knowledge of the redshift evolution of the galaxy bias ($\overline{b_{r}}(z)$) for both the reference and the unknown sample \citep{menard13}. While the clustering amplitude of the unknown sample is not known \emph{a priori}, it can be inferred for the reference population. We determine the luminosity-weighted bias evolution for the reference sample using the fitting formula from \citet{zehavi11}. We present the resultant clustering amplitude on the right of Figure \ref{fig:reffig}. 

For the SDSS photometric galaxies, we adopt an estimate of the unknown sample's bias:
\begin{equation}
\overline{b_u}(z) = \left\{
	\begin{array}{lr}
	1 & : z < 0.1\\
	1 +\,(z - 0.1) & : z \ge 0.1
	\end{array}
\right. \label{eq:bias}
\end{equation}
which corresponds to a flat bias evolution where the unknown sources fully sample the luminosity function, and an increase in the bias as the unknown sources are limited to the most luminous, following the work of \citep{zehavi11}. We compare this estimated bias to that of the reference sample also on the right of Figure \ref{fig:reffig}.


\end{document}